\documentclass[aps,prl,reprint,superscriptaddress,amsfonts,amssymb,amsmath]{revtex4-2}

\usepackage{siunitx}
\usepackage{dcolumn}
\usepackage{graphicx}
\usepackage{dcolumn}
\usepackage{bm}
\usepackage{siunitx}
\usepackage{multirow}
\usepackage[xcdraw]{xcolor}
\usepackage{xcolor}
\usepackage[english]{babel}
\usepackage{array}
\newcolumntype{C}{>{\centering\arraybackslash}p{3em}}
\newcolumntype{E}{>{\centering\arraybackslash}p{5em}}

\begin{document}

\title{Fermi surface and mass renormalization in the iron-based superconductor YFe$_2$Ge$_2$}


\author{Jordan Baglo}
\email[]{Jordan.Baglo@USherbrooke.ca}
\altaffiliation{Now at Institut quantique and D\'epartement de physique, Universit\'e de Sherbrooke, Sherbrooke, Qu\'{e}bec, Canada}
\affiliation{Cavendish Laboratory, University of Cambridge, Cambridge CB3 0HE, UK}

\author{Jiasheng Chen}
\affiliation{Cavendish Laboratory, University of Cambridge, Cambridge CB3 0HE, UK}

\author{Keiron Murphy}
\altaffiliation{Now at Clarendon Laboratory, University of Oxford, Oxford, UK}
\affiliation{Cavendish Laboratory, University of Cambridge, Cambridge CB3 0HE, UK}

\author{Roos Leenen}
\affiliation {High Field Magnet Laboratory (HFML-EMFL), Radboud University, 6525 ED Nijmegen, The Netherlands}

\author{Alix McCollam}
\affiliation {High Field Magnet Laboratory (HFML-EMFL), Radboud University, 6525 ED Nijmegen, The Netherlands}

\author{Michael L. Sutherland}
\affiliation{Cavendish Laboratory, University of Cambridge, Cambridge CB3 0HE, UK}

\author{F.\ Malte Grosche}
\affiliation{Cavendish Laboratory, University of Cambridge, Cambridge CB3 0HE, UK}

\date{\today}

\begin{abstract}
  \noindent 
Quantum oscillation measurements in the new unconventional superconductor YFe$_2$Ge$_2$ resolve all four Fermi surface pockets expected from band structure calculations, which predict an electron pocket in the Brillouin zone corner and three hole pockets enveloping the centers of the top and bottom of the Brillouin zone. The carrier masses are uniformly renormalized by about a factor of five and broadly account for the enhanced heat capacity Sommerfeld coefficient $\simeq \SI{100}{\milli\joule/\mol\square\kelvin}$. Our data highlight the key role of the electron pocket, which despite its small volume accounts for about half the total density of states, and point towards a predominantly local mechanism underlying the mass renormalization in YFe$_2$Ge$_2$.
\end{abstract}

\maketitle
\noindent
Superconductivity in iron pnictides and chalcogenides involves a rich interplay of magnetic, structural, and nematic instabilities as well as the associated quantum critical phenomena \cite{hosono15a,si16,shibauchi14,watson15}. 
This diverse phenomenology includes strong quasiparticle mass renormalization, which in the extreme case of the alkali metal iron arsenides (K/Rb/Cs)Fe$_2$As$_2$ produces heat capacity Sommerfeld coefficients $\sim 100 \relbar\SI{200} {\milli\joule/\mol\square\kelvin}$, approaching values usually associated with rare-earth-based heavy fermion materials. Understanding the origin of this mass renormalization is central to a comprehensive theory of iron-based materials: is it caused by non-local, nearly quantum critical fluctuations associated with incipient order, which also produce the superconducting pairing interaction, or should it be attributed primarily to local, dynamical correlations \cite{tomczak12}, as in the Hund's metal scenario (e.g. \cite{georges13})? 
  
\begin{figure}[t]
\centerline{\includegraphics[width=\columnwidth]{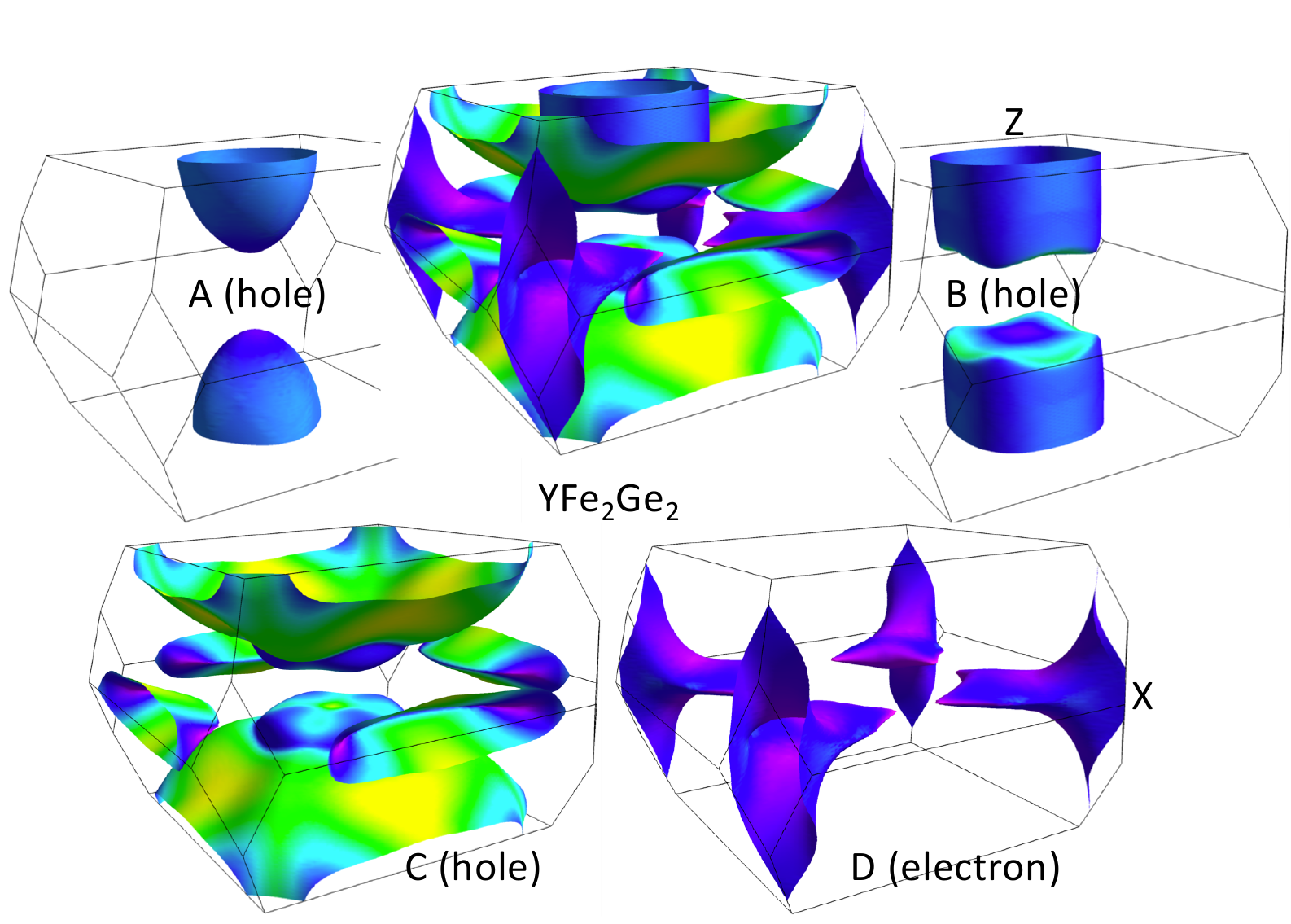}}
\caption{\label{fig:FS} {\em Ab initio} Fermi surface structure of YFe$_2$Ge$_2$. The Fermi surface (top center) consists of three concentric hole
  pockets (A-C), which envelop the Z point on the Brillouin zone (BZ) boundary,
  and a single electron pocket (D) centered on the X point at the corner of the BZ. Lighter colors indicate higher Fermi velocity.}
\end{figure}

The iron germanide superconductor YFe$_2$Ge$_2$ \cite{zou14,chen16,chen19} provides a fresh opportunity to investigate the origin of mass renormalization in strongly correlated iron-based intermetallics. It shares key aspects with the alkali metal iron arsenides (K/Rb/Cs)Fe$_2$As$_2$,
such as (i) bad metal behavior at high temperature $T$, with resistivities $\rho$ of several hundred $\si{\micro\ohm\centi\meter}$, (ii) low transition temperatures $T_c$ of order a few kelvin, (iii) strongly enhanced heat capacity Sommerfeld coefficient $\gamma_n \simeq \SI{100}{\milli\joule/\mol\square\kelvin}$, (iv) reduced heat capacity jump at $T_c$ of order $0.4 T_c \gamma_n$, and residual extrapolated $C/T$ at low $T$ of order $0.4 \gamma_n$ \cite{avila04,hardy13}. On the other hand, it lacks the pnictogen or chalcogen constituents of other Fe-based superconductors, and because the Ge layers in YFe$_2$Ge$_2$ are covalently bonded along $\hat c$, it forms a more compressed, collapsed-tetragonal structure. Density functional theory (DFT) calculations \cite{singh14,subedi14} suggest that this leads to a much more strongly warped Fermi surface (FS) geometry than the cylindrical FS sheets found in other iron-based superconductors (Fig.~\ref{fig:FS}). 

\begin{figure}[t]
\centerline{\includegraphics[width=\columnwidth]{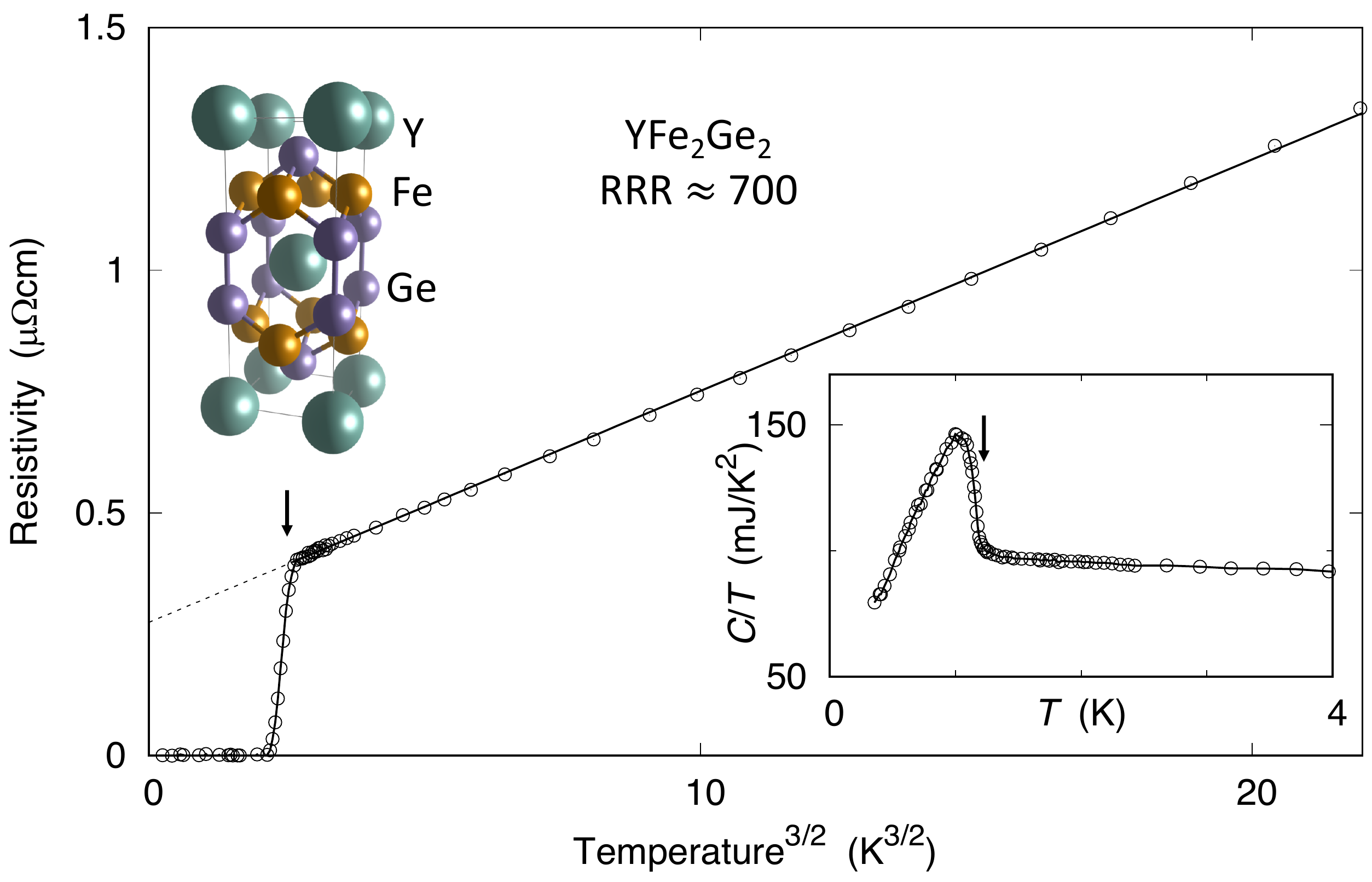}}
\caption{\label{fig:char} Low temperature properties of YFe$_2$Ge$_2$. The in-plane electrical
  resistivity $\rho(T)$ displays a superconducting transition (arrow) below an
  anomalous normal-state $T^{3/2}$
  form. 
  A sharp superconducting anomaly (arrow) is also observed in the heat
  capacity Sommerfeld ratio $C/T$ (inset), which in the normal state
  reaches $\simeq\SI{100}{\milli\joule/\mol\square\kelvin}$. }
\end{figure}

Key input for any comprehensive theoretical description of iron-based superconductors derives from the experimental determination of the electronic structure by photoemission spectroscopy or quantum oscillation measurements (e.g. \cite{coldea08,sato09,terashima13a}). An early ARPES study in YFe$_2$Ge$_2$ \cite{xu16} investigated key aspects of the electronic structure, and core-level spectroscopy indicated a large fluctuating Fe moment $\simeq \SI{1}{\mu_B}$ \cite{sirica15}, which is consistent with neutron scattering and NMR studies \cite{wo19,srpcic17}. The precise determination of Fermi surface and carrier mass by quantum oscillation techniques, however, has  in YFe$_2$Ge$_2$ so far been held back by the lack of single crystals with the required level of purity \cite{kim15a}. 
Here, we present a detailed study of Fermi surface structure and carrier mass by observing de Haas-van Alphen oscillations in a new generation of high-quality crystals of YFe$_2$Ge$_2$. 
We find that the Fermi surface consists of three hole pockets and one electron pocket, with shapes broadly in line with numerical calculations \cite{singh14,subedi14}. Carrier masses are enhanced roughly five-fold over DFT values.
The mass renormalization varies only weakly between Fermi surface pockets and within each Fermi surface pocket, suggesting that on-site interactions provide the primary mechanism for the profound mass enhancement and boost to the
electronic heat capacity recorded in YFe$_2$Ge$_2$.

High-purity crystals of YFe$_2$Ge$_2$ were grown by a liquid transport technique \cite{chen20b,yan17} 
and characterized by electrical transport, magnetic, and thermodynamic measurements. They
display sharp superconducting transition anomalies (Fig. \ref{fig:char}), and their residual resistivities are $\simeq \SI{0.3}{\micro\ohm\cm}$, corresponding to residual resistance ratios $\text{RRR}=\rho(\SI{300}{\kelvin})/\rho_0 \simeq 700$.
Angle-dependent de Haas-van Alphen measurements at fields of up
to $\SI{18}{\tesla}$ were performed using a mutual inductance technique with modulation field amplitude $\simeq \SI{0.2}{\milli\tesla}$ at a
frequency of $\sim\SI{29}{\hertz}$ on two samples (S1, S2)  in a superconducting cryomagnet/dilution
refrigerator system at the Cavendish Laboratory, and using piezoresistive torque magnetometry on a
third sample (S3) at fields of up to $\SI{38}{\tesla}$ in a resistive electromagnet/dilution refrigerator system at
HFML Nijmegen. 
Sample S1 (S2) was mounted with the crystallographic $\hat c$ ($\hat a$) direction aligned with the
axis of the respective pickup coil, and S3 was mounted with  $\hat c$ perpendicular to the surface of the cantilever. 
Quantum oscillation data were extracted by subtracting a low-order polynomial background from the raw data, and
oscillations periodic in $1/B$ were identified from peaks in the power spectrum. The peak frequencies $F$ relate to extremal cross-sectional areas $A_k$ of the Fermi surface via the Onsager relation $A_k = (2\pi e/\hbar) F$. 
The dependence of the signal amplitude $\tilde y$ on temperature $T$ at a fixed magnetic field $B$ provides the effective carrier mass  $m^*$ via the Lifshitz-Kosevich expression
$\tilde y = \alpha T \left[\sinh\left(\SI{14.639}{\tesla\kelvin^{-1}} \frac{T}{B} \frac{m^*}{m_e}\right)\right]^{-1}$,
where $\alpha$ is a temperature-independent factor and $m_e$ is the bare electron mass. The field dependence of the signal envelope provides an estimate of the electronic mean free path \cite{shoenberg09,SuppMat}.
The electronic structure was calculated using the Generalized Gradient
Approximation \cite{perdew96} in WIEN2k \cite{blaha20} with $100,000~ k$-points in the BZ (6768 $k$-points in the irreducible BZ) and $Rk_{max}\!=\!7$, 
using the experimentally determined crystal structure at $\SI{100}{\kelvin}$ with $a=\SI{3.95917(3)}{\angstrom}$, $c=\SI{10.39754(13)}{\angstrom}$, and the fractional vertical Ge position $z=0.378331(7)$ \cite{chen20b}. Extremal orbits, band masses and pocket-resolved contributions to the density of states (DOS) were extracted using SKEAF \cite{rourke12a}, and Fermi surfaces were plotted in FermiSurfer \cite{kawamura19}. 

\begin{figure}
\centerline{\includegraphics[width=\columnwidth]{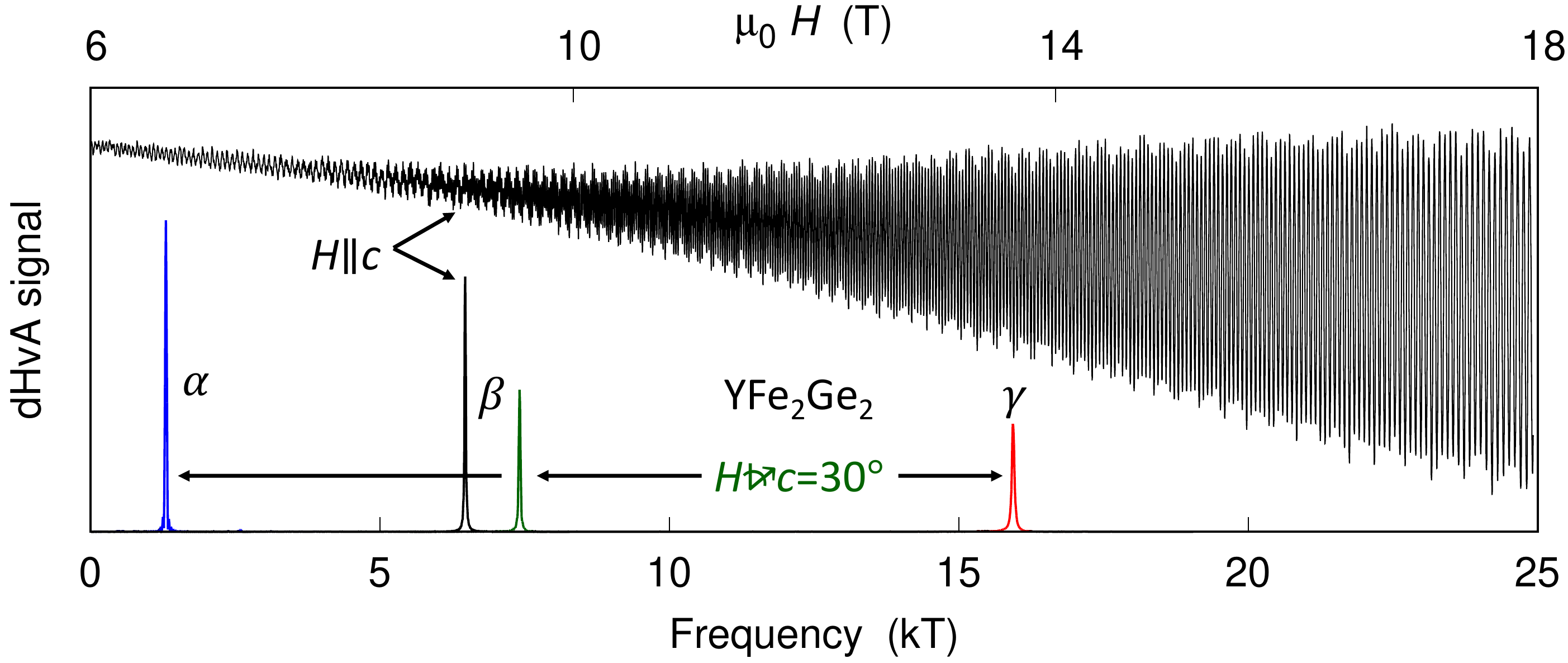}}
\centerline{\includegraphics[width=\columnwidth]{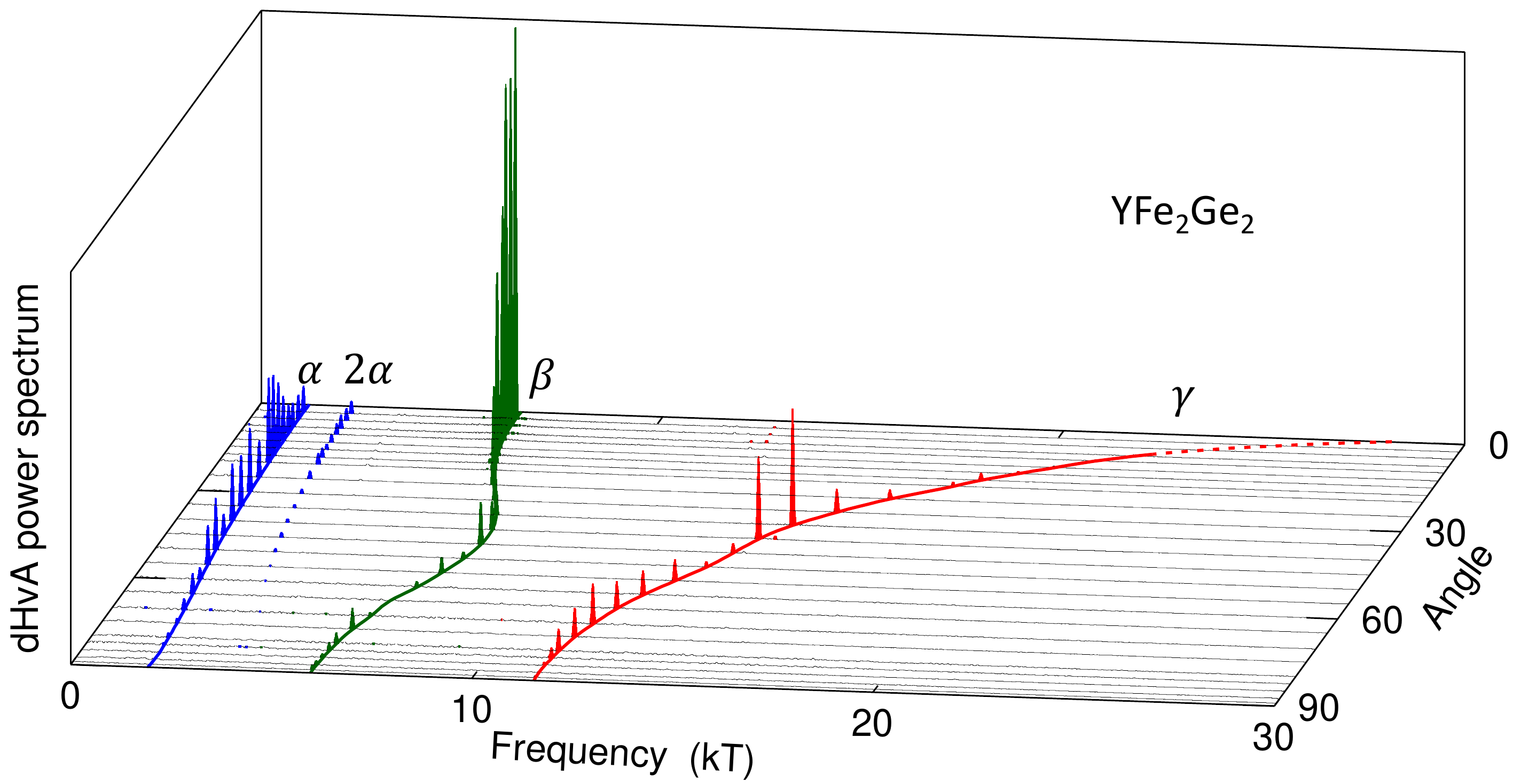}}
\caption{\label{fig:QOtraces} De Haas-van Alphen
  signal: strong oscillations are observed in sample S1 for $B || c$ down to fields $<\SI{6}{\tesla}$ (upper panel, black trace and associated black peak in power spectrum). At a tilt angle $\theta=30^\circ$, three fundamental frequencies $\alpha$, $\beta$, $\gamma$ as well as some harmonics can be resolved. The evolution of these frequencies with tilt angle (lower panel) suggests that they can be assigned to orbits on sheets A, B, and C, respectively (Fig.~\ref{fig:FS}, see text).}
\end{figure}


The calculated Fermi surface (Fig.~\ref{fig:FS}) consists of three roughly ellipsoidal hole pockets (A--C) nested around Z at the
top and bottom of the Brillouin zone and a roughly cylindrical electron pocket (D) around X at the zone corner, with a
duckbill-shaped outgrowth towards the center of the BZ at $\Gamma$. Moreover, our DFT calculations produce an unrenormalized Sommerfeld coefficient $\gamma_0\simeq \SI{16.7}{\milli\joule/\mol\square\kelvin}$, a factor of six less than the experimentally observed value but significantly larger than values previously obtained in calculations based on numerically relaxed $z$ parameters \cite{singh14,subedi14}.

\begin{figure}
\includegraphics[width=\columnwidth]{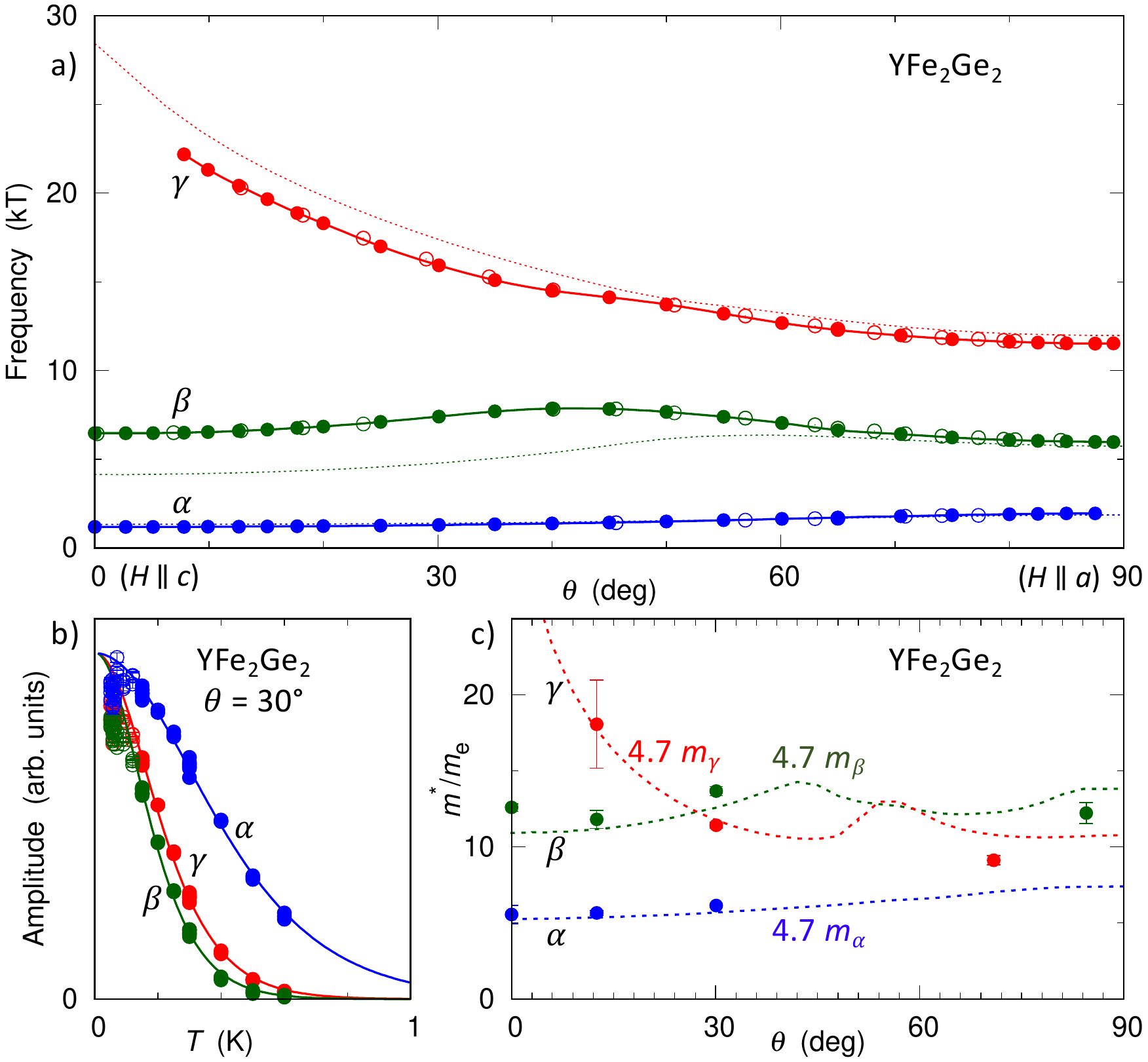}
\caption{\label{fig:rotation} Quantum oscillation results for the three hole pockets. \textbf{(a)} Frequencies (filled symbols: S1, open symbols: S2) as function of angle, rotating from $H\parallel{}\hat{c}$ ($\theta = 0^{\circ}$) to $H\parallel{}\hat{a}$ ($\theta = 90^{\circ}$). Dotted lines show DFT results.  A band shift of $\SI{-10}{mRyd}$ was used for $\alpha$ (Tab.~\ref{tab:budgets}) \textbf{(b)}  Temperature dependence of the oscillation amplitudes extracted in S1 for $\SI{16}{\tesla}<B<\SI{18}{\tesla}$ at $\theta=30^\circ$. Solid lines show Lifshitz-Kosevich fits to the data at $T>\SI{0.12}{\kelvin}$. {\bf (c)} Angle dependence of the extracted effective masses for the $\alpha$, $\beta$, and $\gamma$ orbits compared to DFT band masses $m_{\alpha, \beta, \gamma}$, uniformly renormalized by  $\times 4.7$ (dashed lines).}
\end{figure}

The quantum oscillation signal observed in S1 (Fig.~ \ref{fig:QOtraces}) displays three fundamental frequencies, $\alpha$, $\beta$, and
$\gamma$. The same set of frequencies is present in sample S2  \cite{SuppMat}.
The results of a rotation study from $H\parallel{}\hat{c}$
($\theta = 0^{\circ}$) to $H\parallel{}\hat{a}$ ($\theta = 90^{\circ}$) are summarized in Figs.~\ref{fig:QOtraces}b and \ref{fig:rotation}a. The three fundamental frequencies $\alpha$,
$\beta$ and $\gamma$ were tracked over most of the angular range. 
By comparing to DFT results, they can be assigned to extremal orbits on the three hole pockets. The highest frequency,
$\gamma$, is unambiguously associated with the largest hole pocket, C. The next highest
frequency, $\beta$, matches predictions for the second hole sheet (B), with good quantitative agreement near $H\parallel{}\hat{a}$. The third frequency, $\alpha$, depends weakly on tilt angle, as expected for the smallest, nearly ellipsoidal hole pocket (A), but at
roughly half the predicted frequency.  
The effective masses extracted from Lifshitz-Kosevich fits (Fig.~\ref{fig:rotation}b) are high, with $m^*$ exceeding $10~m_e$ on the $\gamma$ and $\beta$ sheets. These values markedly exceed the highest masses measured in the BaFe$_2$(As/P)$_2$ series \cite{shishido10} and are as high as those observed in (K/Rb/Cs)Fe$_2$As$_2$ \cite{terashima13a,zocco14,eilers16}. Whereas effective masses for $\alpha$ and $\beta$ show little angle dependence, that of $\gamma$ rises sharply near $H\parallel \hat{c}$ (Fig.~\ref{fig:rotation}c). A uniform renormalization of DFT band masses by $\sim 5$ produces good agreement with experimental values on all three hole pockets.


\begin{table}
\begin{ruledtabular}
\begin{tabular}{l*{4}{C}E}
FS pocket & A & B & C & $\sum_{A-C}$ & D \\
Hole count & f.u.$^{-1}$ & ... 	& ...	& ... & ...\\
\hline
{\em Ab initio} & 0.098 & 0.208 &	0.836	& 1.142 & -0.142	\\
QO-corrected & 0.029	& 0.271 &	0.769	& 1.069 & -0.069 \\
\hline
$\Delta E$  (mRyd)&	-10 & 3.5 & -2.4 & & 1.6 \\
\hline \hline
FS pocket & A & B & C & D & $4.7\sum_{A-D}$  \\
$C/T$ &  $\text{mJ}/\text{mol K}^{2}$ &  & ...  & ... & ...  \\
\hline
{\em Ab initio}	& 1.68 &2.95 &5.30 & 6.81 & 78.5 \\ 
QO-corrected & 0.79 & 3.34 & 4.43 & 7.58 & 75.9\\

\end{tabular}
\end{ruledtabular}
\caption{Particle number and DOS budgets in YFe$_2$Ge$_2$. DFT Fermi surface pocket volumes for the hole pockets A--C (top line) have been corrected by using the measured cross-sectional areas (second line, see text). The requirement to reach 1 hole/formula unit (f.u.) fixes the volume of the electron pocket. The third data line lists the band shifts that bring each pocket in line with the QO-corrected hole count. The {\em ab initio} DOS (fourth data line, converted to $C/T$) calculated for the experimentally determined fractional vertical Ge coordinate $z =0.37833$ is dominated by the contribution from the electron pocket D. The analogous calculation for the shifted bands produces a lower contribution from the hole pockets, shifting the balance even more towards the electron pocket. Applying the mass renormalization $\simeq 4.7$ estimated from QO measurements (Fig.~\ref{fig:rotation}) to the calculated DOS produces an overall $C/T$ in rough agreement with experiment. 
}
\label{tab:budgets}

\end{table}

\begin{figure}
\centerline{\includegraphics[width=0.85\columnwidth]{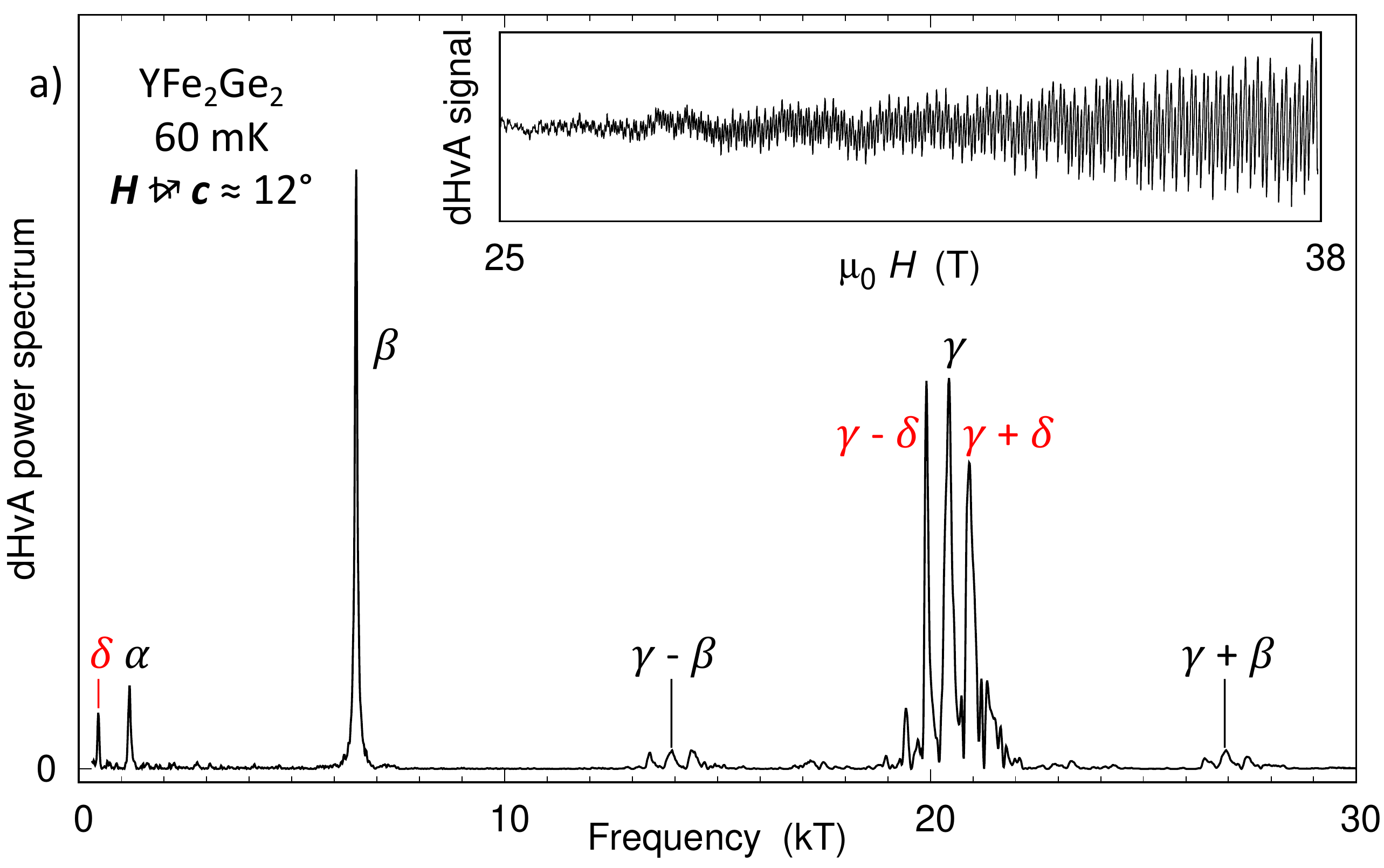}}
\centerline{\includegraphics[width=0.83\columnwidth]{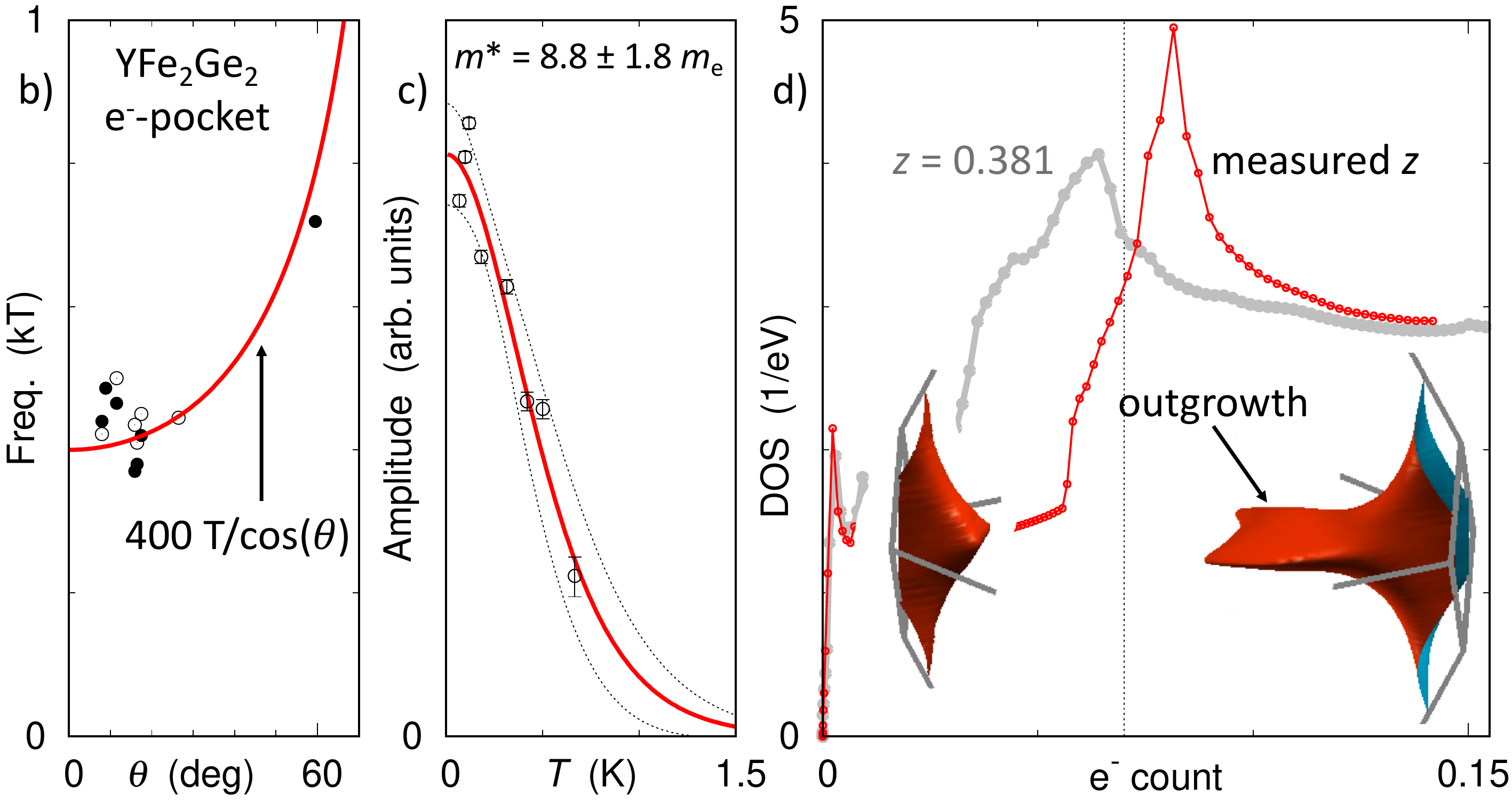}}
\caption{\label{fig:HFML}Electron pocket in YFe$_2$Ge$_2$. {\bf (a)} Quantum oscillations in the magnetic torque occur at the frequencies expected for the three hole pockets $\alpha$, $\beta$, $\gamma$, but the nonlinear response of the cantilever torque sensor produces side lobes to the $\gamma$-frequency which can be indexed as $\gamma \pm \delta$, where $\delta \simeq \SI{450}{\tesla}$ is attributed to the electron pocket located in the BZ corner. A peak at the same frequency $\delta$ is observed at some field angles. {\bf (b)} Tracking $\delta$ via the side-lobe frequency (open symbols) and stand-alone peak frequency (closed symbols) reveals an angle dependence that is consistent with a cylindrical pocket, for which the cross-sectional area is $\propto 1/\cos(\theta)$. {\bf (c)} A Lifshitz-Kosevich fit to the amplitude extracted at an angle of $\simeq 60 ^\circ$ reveals a strongly enhanced carrier mass $m^* = 8.8 \pm 1.8~ m_e$. {\bf (d)} The electron pocket DOS peaks sharply at a critical band filling that depends on $z$, as a duckbill outgrowth appears.}
\end{figure}

The electron pocket D could not be resolved at fields of up to \SI{18}{\tesla}. Its {\em ab initio} volume corresponds to 0.142 electrons/formula unit (f.u.), and its expected dominant QO frequency is $\sim \SI{0.9}{\kilo\tesla}$. This estimate can be refined by using the QO data on the three observed hole pockets in combination with charge neutrality, which imposes an overall hole count of 1/formula unit. Assuming that the {\em ab initio} geometry and associated hole count $n_{0}$ of each hole pocket can be scaled to make them consistent with the measured QO frequencies, we estimate the actual hole count for each pocket as $n = n_{0} q_a^2 q_c / (q_{0,a}^2 q_{0,c} ) \propto \sqrt{F_a^2 F_c}/\sqrt{F_{0,a}^2 F_{0,c}}$. Here, $F_a \propto q_a q_c$, the  QO frequency for field along $\hat{a}$, is given by the extent $q_a, q_c$ of the pocket along $\hat{a}$ and $\hat{c}$, and $F_c \propto q_a^2$, the QO frequency for field along $\hat{c}$, is determined by the extent of the pocket $\perp \hat{c}$, or along $\hat{a}$. $F_{0,a}$ and $F_{0,c}$ are the {\em ab initio} frequencies  for the same pocket and $q_{0,a}$, $q_{0,c}$ the corresponding dimensions.
In the case of pocket C, the $\gamma$ frequency was extrapolated to $\theta=0^\circ$ and the resulting hole count checked for consistency against a second calculation in which $F_c$ and $F_{0,c}$ values were taken at $\theta=7.5^\circ$, the lowest angle at which the $\gamma$ oscillation could be observed. The upper half of Table \ref{tab:budgets} summarizes the results of these calculations, which suggests that the total hole count per f.u. arising from A, B and C is about 1.07. This implies that the volume of the electron pocket is smaller than expected from {\em ab initio} calculations. It corresponds to about $0.07~e^-$ per f.u., which leads to a dominant QO frequency of order $\SI{500}{\tesla}$. 
A clear signature of oscillations in this frequency range was seen in torque magnetometry to fields of up to $\SI{38}{\tesla}$ in sample S3 (Fig.~\ref{fig:HFML}). Our initial observation exploited the phenomenon of torque interaction \cite{vanderkooy68,shoenberg09,SuppMat}, which mixes quantum oscillation frequencies via the nonlinear cantilever response at high magnetic fields. The $\gamma$ peak develops several side lobes, some of which
can be indexed as $\gamma \pm n\delta$ with $\delta \sim \SI{450}{T}$, depending on field angle. At some angles, a distinct low-frequency peak at the corresponding frequency $\delta$ can also be resolved directly (Fig.~\ref{fig:HFML}). Further work will be required to narrow down the current mass estimate $m^* \simeq 8.8\pm 1.8 ~m_e$ obtained at $60^\circ$ tilt angle from the temperature dependence of the directly observed peak intensity (Fig.~\ref{fig:HFML}c), and to track the mass as a function of angle. Because the angle dependence (Fig.~\ref{fig:HFML}b) is consistent with a small cylindrical pocket, we interpret $\delta$ as a signature of the elusive electron pocket D.

The lower part of Table \ref{tab:budgets} tracks the contributions of the different Fermi surface pockets to the overall DOS. When the bands are shifted to bring the DFT pocket volumes into line with QO data, the resulting ``QO-corrected'' DOS is dominated by the electron pocket, which contributes nearly half of the total DOS. Applying a uniform mass renormalization of 4.7 across all pockets A--D, as suggested by the high-resolution measurements for pockets A--C (Fig.~\ref{fig:rotation}) would account for a Sommerfeld coefficient of about $\SI {76}{\milli\joule/\mol\square\kelvin}$. The shortfall of about 20\% compared to the experimental value at zero field  may be attributed to a general reduction of the Sommerfeld coefficient in high magnetic field, as magnetic fluctuations are increasingly suppressed, or we may have underestimated the contribution of the electron pocket, which is highly sensitive to details such as band filling and the structural $z$ parameter. This sensitivity is caused by a flat region in the dispersion along the BZ diagonal, which produces a duckbill outgrowth towards the center of the BZ. As this feature is quasi-1D (little dispersion within the symmetry plane of the BZ), its emergence produces a van Hove-like singularity in the DOS (Fig.~\ref{fig:HFML}d). YFe$_2$Ge$_2$ appears to be situated close to the cusp of this anomaly, similar to the situation of the $\gamma$ sheet in Sr$_2$RuO$_4$ (e.g. \cite{barber18}). This sensitivity introduces significant uncertainty into DFT estimates of the electron pocket DOS contribution.


Our findings establish the electronic structure of YFe$_2$Ge$_2$ as consistent with expectations from DFT calculations, but with strong mass renormalization.
Only less than half of the experimental heat capacity can be attributed to the three hole pockets, for which the mass renormalization of order 5 can be comparatively precisely determined (Tab.~\ref{tab:budgets}). Attributing the missing heat capacity to the electron pocket requires a mass enhancement of a similar magnitude. Torque oscillation data at up to $\SI{38}{\tesla}$ suggest an enhanced carrier mass which is in line with expectations, depending on the precise contribution of the unusual duckbill protrusion of the electron pocket. More detailed measurements will be needed to map out this important pocket fully.
YFe$_2$Ge$_2$ emerges as an unusual iron-based superconductor with a far more three-dimensional electronic structure than the alkali metal iron arsenides (K/Rb/Cs)Fe$_2$As$_2$, with which it shares remarkably similar bulk properties.
The observed uniform mass renormalization is reminiscent of the intensely studied rare-earth-based heavy fermion materials, but 
it occurs without their partially filled $f$ states. This suggests an on-site mechanism like the Kondo effect \cite{wu16} amplified by Hund's rule coupling, which forces the iron $d$ electrons into a high spin state and thereby boosts electronic correlations \cite{haule09,yin11,georges13,demedici14,backes15,zhao20}. 

\begin{acknowledgments}
We thank, in particular, G. Lonzarich for helpful discussions. The work was supported by the EPSRC of the UK (grants no. EP/K012894 and EP/P023290/1), by Trinity College, and by the HFML-RU member of the European Magnetic Field Laboratory (EMFL).
\end{acknowledgments}


\begin{thebibliography}{40}%
\makeatletter
\providecommand \@ifxundefined [1]{%
 \@ifx{#1\undefined}
}%
\providecommand \@ifnum [1]{%
 \ifnum #1\expandafter \@firstoftwo
 \else \expandafter \@secondoftwo
 \fi
}%
\providecommand \@ifx [1]{%
 \ifx #1\expandafter \@firstoftwo
 \else \expandafter \@secondoftwo
 \fi
}%
\providecommand \natexlab [1]{#1}%
\providecommand \enquote  [1]{``#1''}%
\providecommand \bibnamefont  [1]{#1}%
\providecommand \bibfnamefont [1]{#1}%
\providecommand \citenamefont [1]{#1}%
\providecommand \href@noop [0]{\@secondoftwo}%
\providecommand \href [0]{\begingroup \@sanitize@url \@href}%
\providecommand \@href[1]{\@@startlink{#1}\@@href}%
\providecommand \@@href[1]{\endgroup#1\@@endlink}%
\providecommand \@sanitize@url [0]{\catcode `\\12\catcode `\$12\catcode
  `\&12\catcode `\#12\catcode `\^12\catcode `\_12\catcode `\%12\relax}%
\providecommand \@@startlink[1]{}%
\providecommand \@@endlink[0]{}%
\providecommand \url  [0]{\begingroup\@sanitize@url \@url }%
\providecommand \@url [1]{\endgroup\@href {#1}{\urlprefix }}%
\providecommand \urlprefix  [0]{URL }%
\providecommand \Eprint [0]{\href }%
\providecommand \doibase [0]{https://doi.org/}%
\providecommand \selectlanguage [0]{\@gobble}%
\providecommand \bibinfo  [0]{\@secondoftwo}%
\providecommand \bibfield  [0]{\@secondoftwo}%
\providecommand \translation [1]{[#1]}%
\providecommand \BibitemOpen [0]{}%
\providecommand \bibitemStop [0]{}%
\providecommand \bibitemNoStop [0]{.\EOS\space}%
\providecommand \EOS [0]{\spacefactor3000\relax}%
\providecommand \BibitemShut  [1]{\csname bibitem#1\endcsname}%
\let\auto@bib@innerbib\@empty
\bibitem [{\citenamefont {Hosono}\ and\ \citenamefont
  {Kuroki}(2015)}]{hosono15a}%
  \BibitemOpen
  \bibfield  {author} {\bibinfo {author} {\bibfnamefont {H.}~\bibnamefont
  {Hosono}}\ and\ \bibinfo {author} {\bibfnamefont {K.}~\bibnamefont
  {Kuroki}},\ }\href@noop {} {\bibfield  {journal} {\bibinfo  {journal}
  {Physica C: Superconductivity and its Applications}\ }\textbf {\bibinfo
  {volume} {514}},\ \bibinfo {pages} {399} (\bibinfo {year}
  {2015})}\BibitemShut {NoStop}%
\bibitem [{\citenamefont {Si}\ \emph {et~al.}(2016)\citenamefont {Si},
  \citenamefont {Yu},\ and\ \citenamefont {Abrahams}}]{si16}%
  \BibitemOpen
  \bibfield  {author} {\bibinfo {author} {\bibfnamefont {Q.}~\bibnamefont
  {Si}}, \bibinfo {author} {\bibfnamefont {R.}~\bibnamefont {Yu}},\ and\
  \bibinfo {author} {\bibfnamefont {E.}~\bibnamefont {Abrahams}},\ }\href@noop
  {} {\bibfield  {journal} {\bibinfo  {journal} {Nat Rev Mater}\ }\textbf
  {\bibinfo {volume} {1}},\ \bibinfo {pages} {16017} (\bibinfo {year}
  {2016})}\BibitemShut {NoStop}%
\bibitem [{\citenamefont {Shibauchi}\ \emph {et~al.}(2014)\citenamefont
  {Shibauchi}, \citenamefont {Carrington},\ and\ \citenamefont
  {Matsuda}}]{shibauchi14}%
  \BibitemOpen
  \bibfield  {author} {\bibinfo {author} {\bibfnamefont {T.}~\bibnamefont
  {Shibauchi}}, \bibinfo {author} {\bibfnamefont {A.}~\bibnamefont
  {Carrington}},\ and\ \bibinfo {author} {\bibfnamefont {Y.}~\bibnamefont
  {Matsuda}},\ }\href@noop {} {\bibfield  {journal} {\bibinfo  {journal} {Annu.
  Rev. Condens. Matter Phys.}\ }\textbf {\bibinfo {volume} {5}},\ \bibinfo
  {pages} {113} (\bibinfo {year} {2014})}\BibitemShut {NoStop}%
\bibitem [{\citenamefont {Watson}\ \emph {et~al.}(2015)\citenamefont {Watson},
  \citenamefont {Kim}, \citenamefont {Haghighirad}, \citenamefont {Davies},
  \citenamefont {McCollam}, \citenamefont {Narayanan}, \citenamefont {Blake},
  \citenamefont {Chen}, \citenamefont {Ghannadzadeh}, \citenamefont
  {Schofield}, \citenamefont {Hoesch}, \citenamefont {Meingast}, \citenamefont
  {Wolf},\ and\ \citenamefont {Coldea}}]{watson15}%
  \BibitemOpen
  \bibfield  {author} {\bibinfo {author} {\bibfnamefont {M.~D.}\ \bibnamefont
  {Watson}}, \bibinfo {author} {\bibfnamefont {T.~K.}\ \bibnamefont {Kim}},
  \bibinfo {author} {\bibfnamefont {A.~A.}\ \bibnamefont {Haghighirad}},
  \bibinfo {author} {\bibfnamefont {N.~R.}\ \bibnamefont {Davies}}, \bibinfo
  {author} {\bibfnamefont {A.}~\bibnamefont {McCollam}}, \bibinfo {author}
  {\bibfnamefont {A.}~\bibnamefont {Narayanan}}, \bibinfo {author}
  {\bibfnamefont {S.~F.}\ \bibnamefont {Blake}}, \bibinfo {author}
  {\bibfnamefont {Y.~L.}\ \bibnamefont {Chen}}, \bibinfo {author}
  {\bibfnamefont {S.}~\bibnamefont {Ghannadzadeh}}, \bibinfo {author}
  {\bibfnamefont {A.~J.}\ \bibnamefont {Schofield}}, \bibinfo {author}
  {\bibfnamefont {M.}~\bibnamefont {Hoesch}}, \bibinfo {author} {\bibfnamefont
  {C.}~\bibnamefont {Meingast}}, \bibinfo {author} {\bibfnamefont
  {T.}~\bibnamefont {Wolf}},\ and\ \bibinfo {author} {\bibfnamefont {A.~I.}\
  \bibnamefont {Coldea}},\ }\href@noop {} {\bibfield  {journal} {\bibinfo
  {journal} {Phys. Rev. B}\ }\textbf {\bibinfo {volume} {91}},\ \bibinfo
  {pages} {155106} (\bibinfo {year} {2015})}\BibitemShut {NoStop}%
\bibitem [{\citenamefont {Tomczak}\ \emph {et~al.}(2012)\citenamefont
  {Tomczak}, \citenamefont {{van Schilfgaarde}},\ and\ \citenamefont
  {Kotliar}}]{tomczak12}%
  \BibitemOpen
  \bibfield  {author} {\bibinfo {author} {\bibfnamefont {J.~M.}\ \bibnamefont
  {Tomczak}}, \bibinfo {author} {\bibfnamefont {M.}~\bibnamefont {{van
  Schilfgaarde}}},\ and\ \bibinfo {author} {\bibfnamefont {G.}~\bibnamefont
  {Kotliar}},\ }\href@noop {} {\bibfield  {journal} {\bibinfo  {journal} {Phys.
  Rev. Lett.}\ }\textbf {\bibinfo {volume} {109}},\ \bibinfo {pages} {237010}
  (\bibinfo {year} {2012})}\BibitemShut {NoStop}%
\bibitem [{\citenamefont {Georges}\ \emph {et~al.}(2013)\citenamefont
  {Georges}, \citenamefont {de'{\aftergroup\ignorespaces} Medici},\ and\
  \citenamefont {Mravlje}}]{georges13}%
  \BibitemOpen
  \bibfield  {author} {\bibinfo {author} {\bibfnamefont {A.}~\bibnamefont
  {Georges}}, \bibinfo {author} {\bibfnamefont {L.}~\bibnamefont
  {de'{\aftergroup\ignorespaces} Medici}},\ and\ \bibinfo {author}
  {\bibfnamefont {J.}~\bibnamefont {Mravlje}},\ }\href@noop {} {\bibfield
  {journal} {\bibinfo  {journal} {Annu. Rev. Condens. Matter Phys.}\ }\textbf
  {\bibinfo {volume} {4}},\ \bibinfo {pages} {137} (\bibinfo {year}
  {2013})}\BibitemShut {NoStop}%
\bibitem [{\citenamefont {Zou}\ \emph {et~al.}(2014)\citenamefont {Zou},
  \citenamefont {Feng}, \citenamefont {Logg}, \citenamefont {Chen},
  \citenamefont {Lampronti},\ and\ \citenamefont {Grosche}}]{zou14}%
  \BibitemOpen
  \bibfield  {author} {\bibinfo {author} {\bibfnamefont {Y.}~\bibnamefont
  {Zou}}, \bibinfo {author} {\bibfnamefont {Z.}~\bibnamefont {Feng}}, \bibinfo
  {author} {\bibfnamefont {P.~W.}\ \bibnamefont {Logg}}, \bibinfo {author}
  {\bibfnamefont {J.}~\bibnamefont {Chen}}, \bibinfo {author} {\bibfnamefont
  {G.}~\bibnamefont {Lampronti}},\ and\ \bibinfo {author} {\bibfnamefont
  {F.~M.}\ \bibnamefont {Grosche}},\ }\href@noop {} {\bibfield  {journal}
  {\bibinfo  {journal} {Phys. Status Solidi RRL - Rapid Res. Lett.}\ }\textbf
  {\bibinfo {volume} {8}},\ \bibinfo {pages} {928} (\bibinfo {year}
  {2014})}\BibitemShut {NoStop}%
\bibitem [{\citenamefont {Chen}\ \emph {et~al.}(2016)\citenamefont {Chen},
  \citenamefont {Semeniuk}, \citenamefont {Feng}, \citenamefont {Reiss},
  \citenamefont {Brown}, \citenamefont {Zou}, \citenamefont {Logg},
  \citenamefont {Lampronti},\ and\ \citenamefont {Grosche}}]{chen16}%
  \BibitemOpen
  \bibfield  {author} {\bibinfo {author} {\bibfnamefont {J.}~\bibnamefont
  {Chen}}, \bibinfo {author} {\bibfnamefont {K.}~\bibnamefont {Semeniuk}},
  \bibinfo {author} {\bibfnamefont {Z.}~\bibnamefont {Feng}}, \bibinfo {author}
  {\bibfnamefont {P.}~\bibnamefont {Reiss}}, \bibinfo {author} {\bibfnamefont
  {P.}~\bibnamefont {Brown}}, \bibinfo {author} {\bibfnamefont
  {Y.}~\bibnamefont {Zou}}, \bibinfo {author} {\bibfnamefont {P.~W.}\
  \bibnamefont {Logg}}, \bibinfo {author} {\bibfnamefont {G.~I.}\ \bibnamefont
  {Lampronti}},\ and\ \bibinfo {author} {\bibfnamefont {F.~M.}\ \bibnamefont
  {Grosche}},\ }\href@noop {} {\bibfield  {journal} {\bibinfo  {journal} {Phys.
  Rev. Lett.}\ }\textbf {\bibinfo {volume} {116}},\ \bibinfo {pages} {127001}
  (\bibinfo {year} {2016})}\BibitemShut {NoStop}%
\bibitem [{\citenamefont {Chen}\ \emph {et~al.}(2019)\citenamefont {Chen},
  \citenamefont {Gam{\.z}a}, \citenamefont {Semeniuk},\ and\ \citenamefont
  {Grosche}}]{chen19}%
  \BibitemOpen
  \bibfield  {author} {\bibinfo {author} {\bibfnamefont {J.}~\bibnamefont
  {Chen}}, \bibinfo {author} {\bibfnamefont {M.~B.}\ \bibnamefont {Gam{\.z}a}},
  \bibinfo {author} {\bibfnamefont {K.}~\bibnamefont {Semeniuk}},\ and\
  \bibinfo {author} {\bibfnamefont {F.~M.}\ \bibnamefont {Grosche}},\
  }\href@noop {} {\bibfield  {journal} {\bibinfo  {journal} {Phys. Rev. B}\
  }\textbf {\bibinfo {volume} {99}},\ \bibinfo {pages} {020501(R)} (\bibinfo
  {year} {2019})}\BibitemShut {NoStop}%
\bibitem [{\citenamefont {Avila}\ \emph {et~al.}(2004)\citenamefont {Avila},
  \citenamefont {Bud'ko},\ and\ \citenamefont {Canfield}}]{avila04}%
  \BibitemOpen
  \bibfield  {author} {\bibinfo {author} {\bibfnamefont {M.}~\bibnamefont
  {Avila}}, \bibinfo {author} {\bibfnamefont {S.}~\bibnamefont {Bud'ko}},\ and\
  \bibinfo {author} {\bibfnamefont {P.}~\bibnamefont {Canfield}},\ }\href@noop
  {} {\bibfield  {journal} {\bibinfo  {journal} {J. Magn. Magn. Mater.}\
  }\textbf {\bibinfo {volume} {270}},\ \bibinfo {pages} {51} (\bibinfo {year}
  {2004})}\BibitemShut {NoStop}%
\bibitem [{\citenamefont {Hardy}\ \emph {et~al.}(2013)\citenamefont {Hardy},
  \citenamefont {B{\"o}hmer}, \citenamefont {Aoki}, \citenamefont {Burger},
  \citenamefont {Wolf}, \citenamefont {Schweiss}, \citenamefont {Heid},
  \citenamefont {Adelmann}, \citenamefont {Yao}, \citenamefont {Kotliar},
  \citenamefont {Schmalian},\ and\ \citenamefont {Meingast}}]{hardy13}%
  \BibitemOpen
  \bibfield  {author} {\bibinfo {author} {\bibfnamefont {F.}~\bibnamefont
  {Hardy}}, \bibinfo {author} {\bibfnamefont {A.~E.}\ \bibnamefont
  {B{\"o}hmer}}, \bibinfo {author} {\bibfnamefont {D.}~\bibnamefont {Aoki}},
  \bibinfo {author} {\bibfnamefont {P.}~\bibnamefont {Burger}}, \bibinfo
  {author} {\bibfnamefont {T.}~\bibnamefont {Wolf}}, \bibinfo {author}
  {\bibfnamefont {P.}~\bibnamefont {Schweiss}}, \bibinfo {author}
  {\bibfnamefont {R.}~\bibnamefont {Heid}}, \bibinfo {author} {\bibfnamefont
  {P.}~\bibnamefont {Adelmann}}, \bibinfo {author} {\bibfnamefont {Y.~X.}\
  \bibnamefont {Yao}}, \bibinfo {author} {\bibfnamefont {G.}~\bibnamefont
  {Kotliar}}, \bibinfo {author} {\bibfnamefont {J.}~\bibnamefont {Schmalian}},\
  and\ \bibinfo {author} {\bibfnamefont {C.}~\bibnamefont {Meingast}},\
  }\href@noop {} {\bibfield  {journal} {\bibinfo  {journal} {Phys. Rev. Lett.}\
  }\textbf {\bibinfo {volume} {111}},\ \bibinfo {pages} {027002} (\bibinfo
  {year} {2013})}\BibitemShut {NoStop}%
\bibitem [{\citenamefont {Singh}(2014)}]{singh14}%
  \BibitemOpen
  \bibfield  {author} {\bibinfo {author} {\bibfnamefont {D.~J.}\ \bibnamefont
  {Singh}},\ }\href@noop {} {\bibfield  {journal} {\bibinfo  {journal} {Phys.
  Rev. B}\ }\textbf {\bibinfo {volume} {89}},\ \bibinfo {pages} {024505}
  (\bibinfo {year} {2014})}\BibitemShut {NoStop}%
\bibitem [{\citenamefont {Subedi}(2014)}]{subedi14}%
  \BibitemOpen
  \bibfield  {author} {\bibinfo {author} {\bibfnamefont {A.}~\bibnamefont
  {Subedi}},\ }\href@noop {} {\bibfield  {journal} {\bibinfo  {journal} {Phys.
  Rev. B}\ }\textbf {\bibinfo {volume} {89}},\ \bibinfo {pages} {024504}
  (\bibinfo {year} {2014})}\BibitemShut {NoStop}%
\bibitem [{\citenamefont {Coldea}\ \emph {et~al.}(2008)\citenamefont {Coldea},
  \citenamefont {Fletcher}, \citenamefont {Carrington}, \citenamefont
  {Analytis}, \citenamefont {Bangura}, \citenamefont {Chu}, \citenamefont
  {Erickson}, \citenamefont {Fisher}, \citenamefont {Hussey},\ and\
  \citenamefont {McDonald}}]{coldea08}%
  \BibitemOpen
  \bibfield  {author} {\bibinfo {author} {\bibfnamefont {A.~I.}\ \bibnamefont
  {Coldea}}, \bibinfo {author} {\bibfnamefont {J.~D.}\ \bibnamefont
  {Fletcher}}, \bibinfo {author} {\bibfnamefont {A.}~\bibnamefont
  {Carrington}}, \bibinfo {author} {\bibfnamefont {J.~G.}\ \bibnamefont
  {Analytis}}, \bibinfo {author} {\bibfnamefont {A.~F.}\ \bibnamefont
  {Bangura}}, \bibinfo {author} {\bibfnamefont {J.-H.}\ \bibnamefont {Chu}},
  \bibinfo {author} {\bibfnamefont {A.~S.}\ \bibnamefont {Erickson}}, \bibinfo
  {author} {\bibfnamefont {I.~R.}\ \bibnamefont {Fisher}}, \bibinfo {author}
  {\bibfnamefont {N.~E.}\ \bibnamefont {Hussey}},\ and\ \bibinfo {author}
  {\bibfnamefont {R.~D.}\ \bibnamefont {McDonald}},\ }\href@noop {} {\bibfield
  {journal} {\bibinfo  {journal} {Phys. Rev. Lett.}\ }\textbf {\bibinfo
  {volume} {101}},\ \bibinfo {pages} {216402} (\bibinfo {year}
  {2008})}\BibitemShut {NoStop}%
\bibitem [{\citenamefont {Sato}\ \emph {et~al.}(2009)\citenamefont {Sato},
  \citenamefont {Nakayama}, \citenamefont {Sekiba}, \citenamefont {Richard},
  \citenamefont {Xu}, \citenamefont {Souma}, \citenamefont {Takahashi},
  \citenamefont {Chen}, \citenamefont {Luo}, \citenamefont {Wang},\ and\
  \citenamefont {Ding}}]{sato09}%
  \BibitemOpen
  \bibfield  {author} {\bibinfo {author} {\bibfnamefont {T.}~\bibnamefont
  {Sato}}, \bibinfo {author} {\bibfnamefont {K.}~\bibnamefont {Nakayama}},
  \bibinfo {author} {\bibfnamefont {Y.}~\bibnamefont {Sekiba}}, \bibinfo
  {author} {\bibfnamefont {P.}~\bibnamefont {Richard}}, \bibinfo {author}
  {\bibfnamefont {Y.-M.}\ \bibnamefont {Xu}}, \bibinfo {author} {\bibfnamefont
  {S.}~\bibnamefont {Souma}}, \bibinfo {author} {\bibfnamefont
  {T.}~\bibnamefont {Takahashi}}, \bibinfo {author} {\bibfnamefont {G.~F.}\
  \bibnamefont {Chen}}, \bibinfo {author} {\bibfnamefont {J.~L.}\ \bibnamefont
  {Luo}}, \bibinfo {author} {\bibfnamefont {N.~L.}\ \bibnamefont {Wang}},\ and\
  \bibinfo {author} {\bibfnamefont {H.}~\bibnamefont {Ding}},\ }\href@noop {}
  {\bibfield  {journal} {\bibinfo  {journal} {Phys. Rev. Lett.}\ }\textbf
  {\bibinfo {volume} {103}},\ \bibinfo {pages} {047002} (\bibinfo {year}
  {2009})}\BibitemShut {NoStop}%
\bibitem [{\citenamefont {Terashima}\ \emph {et~al.}(2013)\citenamefont
  {Terashima}, \citenamefont {Kurita}, \citenamefont {Kimata}, \citenamefont
  {Tomita}, \citenamefont {Tsuchiya}, \citenamefont {Imai}, \citenamefont
  {Sato}, \citenamefont {Kihou}, \citenamefont {Lee}, \citenamefont {Kito},
  \citenamefont {Eisaki}, \citenamefont {Iyo}, \citenamefont {Saito},
  \citenamefont {Fukazawa}, \citenamefont {Kohori}, \citenamefont {Harima},\
  and\ \citenamefont {Uji}}]{terashima13a}%
  \BibitemOpen
  \bibfield  {author} {\bibinfo {author} {\bibfnamefont {T.}~\bibnamefont
  {Terashima}}, \bibinfo {author} {\bibfnamefont {N.}~\bibnamefont {Kurita}},
  \bibinfo {author} {\bibfnamefont {M.}~\bibnamefont {Kimata}}, \bibinfo
  {author} {\bibfnamefont {M.}~\bibnamefont {Tomita}}, \bibinfo {author}
  {\bibfnamefont {S.}~\bibnamefont {Tsuchiya}}, \bibinfo {author}
  {\bibfnamefont {M.}~\bibnamefont {Imai}}, \bibinfo {author} {\bibfnamefont
  {A.}~\bibnamefont {Sato}}, \bibinfo {author} {\bibfnamefont {K.}~\bibnamefont
  {Kihou}}, \bibinfo {author} {\bibfnamefont {C.-H.}\ \bibnamefont {Lee}},
  \bibinfo {author} {\bibfnamefont {H.}~\bibnamefont {Kito}}, \bibinfo {author}
  {\bibfnamefont {H.}~\bibnamefont {Eisaki}}, \bibinfo {author} {\bibfnamefont
  {A.}~\bibnamefont {Iyo}}, \bibinfo {author} {\bibfnamefont {T.}~\bibnamefont
  {Saito}}, \bibinfo {author} {\bibfnamefont {H.}~\bibnamefont {Fukazawa}},
  \bibinfo {author} {\bibfnamefont {Y.}~\bibnamefont {Kohori}}, \bibinfo
  {author} {\bibfnamefont {H.}~\bibnamefont {Harima}},\ and\ \bibinfo {author}
  {\bibfnamefont {S.}~\bibnamefont {Uji}},\ }\href@noop {} {\bibfield
  {journal} {\bibinfo  {journal} {Phys. Rev. B}\ }\textbf {\bibinfo {volume}
  {87}},\ \bibinfo {pages} {224512} (\bibinfo {year} {2013})}\BibitemShut
  {NoStop}%
\bibitem [{\citenamefont {Xu}\ \emph {et~al.}(2016)\citenamefont {Xu},
  \citenamefont {Shen}, \citenamefont {Zhu}, \citenamefont {Jiang},
  \citenamefont {Xie}, \citenamefont {Wang}, \citenamefont {Pan}, \citenamefont
  {Dudin}, \citenamefont {Kim}, \citenamefont {Hoesch}, \citenamefont {Zhao},
  \citenamefont {Wan},\ and\ \citenamefont {Feng}}]{xu16}%
  \BibitemOpen
  \bibfield  {author} {\bibinfo {author} {\bibfnamefont {D.~F.}\ \bibnamefont
  {Xu}}, \bibinfo {author} {\bibfnamefont {D.~W.}\ \bibnamefont {Shen}},
  \bibinfo {author} {\bibfnamefont {D.}~\bibnamefont {Zhu}}, \bibinfo {author}
  {\bibfnamefont {J.}~\bibnamefont {Jiang}}, \bibinfo {author} {\bibfnamefont
  {B.~P.}\ \bibnamefont {Xie}}, \bibinfo {author} {\bibfnamefont {Q.~S.}\
  \bibnamefont {Wang}}, \bibinfo {author} {\bibfnamefont {B.~Y.}\ \bibnamefont
  {Pan}}, \bibinfo {author} {\bibfnamefont {P.}~\bibnamefont {Dudin}}, \bibinfo
  {author} {\bibfnamefont {T.~K.}\ \bibnamefont {Kim}}, \bibinfo {author}
  {\bibfnamefont {M.}~\bibnamefont {Hoesch}}, \bibinfo {author} {\bibfnamefont
  {J.}~\bibnamefont {Zhao}}, \bibinfo {author} {\bibfnamefont {X.~G.}\
  \bibnamefont {Wan}},\ and\ \bibinfo {author} {\bibfnamefont {D.~L.}\
  \bibnamefont {Feng}},\ }\href@noop {} {\bibfield  {journal} {\bibinfo
  {journal} {Phys. Rev. B}\ }\textbf {\bibinfo {volume} {93}},\ \bibinfo
  {pages} {024506} (\bibinfo {year} {2016})}\BibitemShut {NoStop}%
\bibitem [{\citenamefont {Sirica}\ \emph {et~al.}(2015)\citenamefont {Sirica},
  \citenamefont {Bondino}, \citenamefont {Nappini}, \citenamefont {P{\'i}{\v
  s}}, \citenamefont {Poudel}, \citenamefont {Christianson}, \citenamefont
  {Mandrus}, \citenamefont {Singh},\ and\ \citenamefont {Mannella}}]{sirica15}%
  \BibitemOpen
  \bibfield  {author} {\bibinfo {author} {\bibfnamefont {N.}~\bibnamefont
  {Sirica}}, \bibinfo {author} {\bibfnamefont {F.}~\bibnamefont {Bondino}},
  \bibinfo {author} {\bibfnamefont {S.}~\bibnamefont {Nappini}}, \bibinfo
  {author} {\bibfnamefont {I.}~\bibnamefont {P{\'i}{\v s}}}, \bibinfo {author}
  {\bibfnamefont {L.}~\bibnamefont {Poudel}}, \bibinfo {author} {\bibfnamefont
  {A.~D.}\ \bibnamefont {Christianson}}, \bibinfo {author} {\bibfnamefont
  {D.}~\bibnamefont {Mandrus}}, \bibinfo {author} {\bibfnamefont {D.~J.}\
  \bibnamefont {Singh}},\ and\ \bibinfo {author} {\bibfnamefont
  {N.}~\bibnamefont {Mannella}},\ }\href@noop {} {\bibfield  {journal}
  {\bibinfo  {journal} {Phys. Rev. B}\ }\textbf {\bibinfo {volume} {91}},\
  \bibinfo {pages} {121102(R)} (\bibinfo {year} {2015})}\BibitemShut {NoStop}%
\bibitem [{\citenamefont {Wo}\ \emph {et~al.}(2019)\citenamefont {Wo},
  \citenamefont {Wang}, \citenamefont {Shen}, \citenamefont {Zhang},
  \citenamefont {Hao}, \citenamefont {Feng}, \citenamefont {Shen},
  \citenamefont {He}, \citenamefont {Pan}, \citenamefont {Wang}, \citenamefont
  {Nakajima}, \citenamefont {{Ohira-Kawamura}}, \citenamefont {Steffens},
  \citenamefont {Boehm}, \citenamefont {Schmalzl}, \citenamefont {Forrest},
  \citenamefont {Matsuda}, \citenamefont {Zhao}, \citenamefont {Lynn},
  \citenamefont {Yin},\ and\ \citenamefont {Zhao}}]{wo19}%
  \BibitemOpen
  \bibfield  {author} {\bibinfo {author} {\bibfnamefont {H.}~\bibnamefont
  {Wo}}, \bibinfo {author} {\bibfnamefont {Q.}~\bibnamefont {Wang}}, \bibinfo
  {author} {\bibfnamefont {Y.}~\bibnamefont {Shen}}, \bibinfo {author}
  {\bibfnamefont {X.}~\bibnamefont {Zhang}}, \bibinfo {author} {\bibfnamefont
  {Y.}~\bibnamefont {Hao}}, \bibinfo {author} {\bibfnamefont {Y.}~\bibnamefont
  {Feng}}, \bibinfo {author} {\bibfnamefont {S.}~\bibnamefont {Shen}}, \bibinfo
  {author} {\bibfnamefont {Z.}~\bibnamefont {He}}, \bibinfo {author}
  {\bibfnamefont {B.}~\bibnamefont {Pan}}, \bibinfo {author} {\bibfnamefont
  {W.}~\bibnamefont {Wang}}, \bibinfo {author} {\bibfnamefont {K.}~\bibnamefont
  {Nakajima}}, \bibinfo {author} {\bibfnamefont {S.}~\bibnamefont
  {{Ohira-Kawamura}}}, \bibinfo {author} {\bibfnamefont {P.}~\bibnamefont
  {Steffens}}, \bibinfo {author} {\bibfnamefont {M.}~\bibnamefont {Boehm}},
  \bibinfo {author} {\bibfnamefont {K.}~\bibnamefont {Schmalzl}}, \bibinfo
  {author} {\bibfnamefont {T.~R.}\ \bibnamefont {Forrest}}, \bibinfo {author}
  {\bibfnamefont {M.}~\bibnamefont {Matsuda}}, \bibinfo {author} {\bibfnamefont
  {Y.}~\bibnamefont {Zhao}}, \bibinfo {author} {\bibfnamefont {J.~W.}\
  \bibnamefont {Lynn}}, \bibinfo {author} {\bibfnamefont {Z.}~\bibnamefont
  {Yin}},\ and\ \bibinfo {author} {\bibfnamefont {J.}~\bibnamefont {Zhao}},\
  }\href@noop {} {\bibfield  {journal} {\bibinfo  {journal} {Phys. Rev. Lett.}\
  }\textbf {\bibinfo {volume} {122}},\ \bibinfo {pages} {217003} (\bibinfo
  {year} {2019})}\BibitemShut {NoStop}%
\bibitem [{\citenamefont {Srp{\v c}i{\v c}}\ \emph {et~al.}(2017)\citenamefont
  {Srp{\v c}i{\v c}}, \citenamefont {Jegli{\v c}}, \citenamefont {Felner},
  \citenamefont {Lv}, \citenamefont {Chu},\ and\ \citenamefont {Ar{\v
  c}on}}]{srpcic17}%
  \BibitemOpen
  \bibfield  {author} {\bibinfo {author} {\bibfnamefont {J.}~\bibnamefont
  {Srp{\v c}i{\v c}}}, \bibinfo {author} {\bibfnamefont {P.}~\bibnamefont
  {Jegli{\v c}}}, \bibinfo {author} {\bibfnamefont {I.}~\bibnamefont {Felner}},
  \bibinfo {author} {\bibfnamefont {B.}~\bibnamefont {Lv}}, \bibinfo {author}
  {\bibfnamefont {C.~W.}\ \bibnamefont {Chu}},\ and\ \bibinfo {author}
  {\bibfnamefont {D.}~\bibnamefont {Ar{\v c}on}},\ }\href@noop {} {\bibfield
  {journal} {\bibinfo  {journal} {Phys. Rev. B}\ }\textbf {\bibinfo {volume}
  {96}},\ \bibinfo {pages} {174430} (\bibinfo {year} {2017})}\BibitemShut
  {NoStop}%
\bibitem [{\citenamefont {Kim}\ \emph {et~al.}(2015)\citenamefont {Kim},
  \citenamefont {Ran}, \citenamefont {Mun}, \citenamefont {Hodovanets},
  \citenamefont {Tanatar}, \citenamefont {Prozorov}, \citenamefont {Bud'ko},\
  and\ \citenamefont {Canfield}}]{kim15a}%
  \BibitemOpen
  \bibfield  {author} {\bibinfo {author} {\bibfnamefont {H.}~\bibnamefont
  {Kim}}, \bibinfo {author} {\bibfnamefont {S.}~\bibnamefont {Ran}}, \bibinfo
  {author} {\bibfnamefont {E.}~\bibnamefont {Mun}}, \bibinfo {author}
  {\bibfnamefont {H.}~\bibnamefont {Hodovanets}}, \bibinfo {author}
  {\bibfnamefont {M.}~\bibnamefont {Tanatar}}, \bibinfo {author} {\bibfnamefont
  {R.}~\bibnamefont {Prozorov}}, \bibinfo {author} {\bibfnamefont
  {S.}~\bibnamefont {Bud'ko}},\ and\ \bibinfo {author} {\bibfnamefont
  {P.}~\bibnamefont {Canfield}},\ }\href@noop {} {\bibfield  {journal}
  {\bibinfo  {journal} {Philos. Mag.}\ }\textbf {\bibinfo {volume} {95}},\
  \bibinfo {pages} {804} (\bibinfo {year} {2015})}\BibitemShut {NoStop}%
\bibitem [{\citenamefont {Chen}\ \emph {et~al.}(2020)\citenamefont {Chen},
  \citenamefont {Gam{\.z}a}, \citenamefont {Banda}, \citenamefont {Murphy},
  \citenamefont {Tarrant}, \citenamefont {Brando},\ and\ \citenamefont
  {Grosche}}]{chen20b}%
  \BibitemOpen
  \bibfield  {author} {\bibinfo {author} {\bibfnamefont {J.}~\bibnamefont
  {Chen}}, \bibinfo {author} {\bibfnamefont {M.~B.}\ \bibnamefont {Gam{\.z}a}},
  \bibinfo {author} {\bibfnamefont {J.}~\bibnamefont {Banda}}, \bibinfo
  {author} {\bibfnamefont {K.}~\bibnamefont {Murphy}}, \bibinfo {author}
  {\bibfnamefont {J.}~\bibnamefont {Tarrant}}, \bibinfo {author} {\bibfnamefont
  {M.}~\bibnamefont {Brando}},\ and\ \bibinfo {author} {\bibfnamefont {F.~M.}\
  \bibnamefont {Grosche}},\ }\href@noop {} {\bibfield  {journal} {\bibinfo
  {journal} {Phys. Rev. Lett.}\ }\textbf {\bibinfo {volume} {125}},\ \bibinfo
  {pages} {237002} (\bibinfo {year} {2020})}\BibitemShut {NoStop}%
\bibitem [{\citenamefont {Yan}\ \emph {et~al.}(2017)\citenamefont {Yan},
  \citenamefont {Sales}, \citenamefont {Susner},\ and\ \citenamefont
  {Mcguire}}]{yan17}%
  \BibitemOpen
  \bibfield  {author} {\bibinfo {author} {\bibfnamefont {J.-Q.}\ \bibnamefont
  {Yan}}, \bibinfo {author} {\bibfnamefont {B.~C.}\ \bibnamefont {Sales}},
  \bibinfo {author} {\bibfnamefont {M.~A.}\ \bibnamefont {Susner}},\ and\
  \bibinfo {author} {\bibfnamefont {M.~A.}\ \bibnamefont {Mcguire}},\
  }\href@noop {} {\bibfield  {journal} {\bibinfo  {journal} {Phys. Rev.
  Mater.}\ }\textbf {\bibinfo {volume} {1}},\ \bibinfo {pages} {023402}
  (\bibinfo {year} {2017})}\BibitemShut {NoStop}%
\bibitem [{\citenamefont {Shoenberg}(2009)}]{shoenberg09}%
  \BibitemOpen
  \bibfield  {author} {\bibinfo {author} {\bibfnamefont {D.}~\bibnamefont
  {Shoenberg}},\ }\href@noop {} {\emph {\bibinfo {title} {Magnetic Oscillations
  in Metals}}},\ \bibinfo {edition} {1st}\ ed.\ (\bibinfo  {publisher}
  {Cambridge University Press},\ \bibinfo {address} {{Cambridge}},\ \bibinfo
  {year} {2009})\BibitemShut {NoStop}%
\bibitem [{Sup(2021)}]{SuppMat}%
  \BibitemOpen
  \bibfield  {title} {\bibinfo {title} {{See Supplemental Material at
  [\emph{URL}], which includes Refs. \cite{vanderkooy68,shoenberg09}, for
  further details of QO results in sample S2, the Dingle analysis of the mean
  free path for both samples S1 and S2, and the torque interaction
  phenomenon.}}} (\bibinfo {year} {2021})\BibitemShut {NoStop}%
\bibitem [{\citenamefont {Perdew}\ \emph {et~al.}(1996)\citenamefont {Perdew},
  \citenamefont {Burke},\ and\ \citenamefont {Ernzerhof}}]{perdew96}%
  \BibitemOpen
  \bibfield  {author} {\bibinfo {author} {\bibfnamefont {J.~P.}\ \bibnamefont
  {Perdew}}, \bibinfo {author} {\bibfnamefont {K.}~\bibnamefont {Burke}},\ and\
  \bibinfo {author} {\bibfnamefont {M.}~\bibnamefont {Ernzerhof}},\ }\href@noop
  {} {\bibfield  {journal} {\bibinfo  {journal} {Phys. Rev. Lett.}\ }\textbf
  {\bibinfo {volume} {77}},\ \bibinfo {pages} {3865} (\bibinfo {year}
  {1996})}\BibitemShut {NoStop}%
\bibitem [{\citenamefont {Blaha}\ \emph {et~al.}(2020)\citenamefont {Blaha},
  \citenamefont {Schwarz}, \citenamefont {Tran}, \citenamefont {Laskowski},
  \citenamefont {Madsen},\ and\ \citenamefont {Marks}}]{blaha20}%
  \BibitemOpen
  \bibfield  {author} {\bibinfo {author} {\bibfnamefont {P.}~\bibnamefont
  {Blaha}}, \bibinfo {author} {\bibfnamefont {K.}~\bibnamefont {Schwarz}},
  \bibinfo {author} {\bibfnamefont {F.}~\bibnamefont {Tran}}, \bibinfo {author}
  {\bibfnamefont {R.}~\bibnamefont {Laskowski}}, \bibinfo {author}
  {\bibfnamefont {G.~K.~H.}\ \bibnamefont {Madsen}},\ and\ \bibinfo {author}
  {\bibfnamefont {L.~D.}\ \bibnamefont {Marks}},\ }\href@noop {} {\bibfield
  {journal} {\bibinfo  {journal} {J. Chem. Phys.}\ }\textbf {\bibinfo {volume}
  {152}},\ \bibinfo {pages} {074101} (\bibinfo {year} {2020})}\BibitemShut
  {NoStop}%
\bibitem [{\citenamefont {Rourke}\ and\ \citenamefont
  {Julian}(2012)}]{rourke12a}%
  \BibitemOpen
  \bibfield  {author} {\bibinfo {author} {\bibfnamefont {P.}~\bibnamefont
  {Rourke}}\ and\ \bibinfo {author} {\bibfnamefont {S.}~\bibnamefont
  {Julian}},\ }\href@noop {} {\bibfield  {journal} {\bibinfo  {journal}
  {Computer Physics Communications}\ }\textbf {\bibinfo {volume} {183}},\
  \bibinfo {pages} {324} (\bibinfo {year} {2012})}\BibitemShut {NoStop}%
\bibitem [{\citenamefont {Kawamura}(2019)}]{kawamura19}%
  \BibitemOpen
  \bibfield  {author} {\bibinfo {author} {\bibfnamefont {M.}~\bibnamefont
  {Kawamura}},\ }\href@noop {} {\bibfield  {journal} {\bibinfo  {journal}
  {Computer Physics Communications}\ }\textbf {\bibinfo {volume} {239}},\
  \bibinfo {pages} {197} (\bibinfo {year} {2019})}\BibitemShut {NoStop}%
\bibitem [{\citenamefont {Shishido}\ \emph {et~al.}(2010)\citenamefont
  {Shishido}, \citenamefont {Bangura}, \citenamefont {Coldea}, \citenamefont
  {Tonegawa}, \citenamefont {Hashimoto}, \citenamefont {Kasahara},
  \citenamefont {Rourke}, \citenamefont {Ikeda}, \citenamefont {Terashima},
  \citenamefont {Settai}, \citenamefont {{\=O}nuki}, \citenamefont {Vignolles},
  \citenamefont {Proust}, \citenamefont {Vignolle}, \citenamefont {McCollam},
  \citenamefont {Matsuda}, \citenamefont {Shibauchi},\ and\ \citenamefont
  {Carrington}}]{shishido10}%
  \BibitemOpen
  \bibfield  {author} {\bibinfo {author} {\bibfnamefont {H.}~\bibnamefont
  {Shishido}}, \bibinfo {author} {\bibfnamefont {A.~F.}\ \bibnamefont
  {Bangura}}, \bibinfo {author} {\bibfnamefont {A.~I.}\ \bibnamefont {Coldea}},
  \bibinfo {author} {\bibfnamefont {S.}~\bibnamefont {Tonegawa}}, \bibinfo
  {author} {\bibfnamefont {K.}~\bibnamefont {Hashimoto}}, \bibinfo {author}
  {\bibfnamefont {S.}~\bibnamefont {Kasahara}}, \bibinfo {author}
  {\bibfnamefont {P.~M.~C.}\ \bibnamefont {Rourke}}, \bibinfo {author}
  {\bibfnamefont {H.}~\bibnamefont {Ikeda}}, \bibinfo {author} {\bibfnamefont
  {T.}~\bibnamefont {Terashima}}, \bibinfo {author} {\bibfnamefont
  {R.}~\bibnamefont {Settai}}, \bibinfo {author} {\bibfnamefont
  {Y.}~\bibnamefont {{\=O}nuki}}, \bibinfo {author} {\bibfnamefont
  {D.}~\bibnamefont {Vignolles}}, \bibinfo {author} {\bibfnamefont
  {C.}~\bibnamefont {Proust}}, \bibinfo {author} {\bibfnamefont
  {B.}~\bibnamefont {Vignolle}}, \bibinfo {author} {\bibfnamefont
  {A.}~\bibnamefont {McCollam}}, \bibinfo {author} {\bibfnamefont
  {Y.}~\bibnamefont {Matsuda}}, \bibinfo {author} {\bibfnamefont
  {T.}~\bibnamefont {Shibauchi}},\ and\ \bibinfo {author} {\bibfnamefont
  {A.}~\bibnamefont {Carrington}},\ }\href@noop {} {\bibfield  {journal}
  {\bibinfo  {journal} {Phys. Rev. Lett.}\ }\textbf {\bibinfo {volume} {104}},\
  \bibinfo {pages} {057008} (\bibinfo {year} {2010})}\BibitemShut {NoStop}%
\bibitem [{\citenamefont {Zocco}\ \emph {et~al.}(2014)\citenamefont {Zocco},
  \citenamefont {Grube}, \citenamefont {Eilers}, \citenamefont {Wolf},\ and\
  \citenamefont {L\"ohneysen}}]{zocco14}%
  \BibitemOpen
  \bibfield  {author} {\bibinfo {author} {\bibfnamefont {D.~A.}\ \bibnamefont
  {Zocco}}, \bibinfo {author} {\bibfnamefont {K.}~\bibnamefont {Grube}},
  \bibinfo {author} {\bibfnamefont {F.}~\bibnamefont {Eilers}}, \bibinfo
  {author} {\bibfnamefont {T.}~\bibnamefont {Wolf}},\ and\ \bibinfo {author}
  {\bibfnamefont {H.~v.}\ \bibnamefont {L\"ohneysen}},\ }\href
  {https://doi.org/10.7566/JPSCP.3.015007} {\bibfield  {journal} {\bibinfo
  {journal} {JPS Conf. Proc.}\ }\textbf {\bibinfo {volume} {3}},\ \bibinfo
  {pages} {015007} (\bibinfo {year} {2014})}\BibitemShut {NoStop}%
\bibitem [{\citenamefont {Eilers}\ \emph {et~al.}(2016)\citenamefont {Eilers},
  \citenamefont {Grube}, \citenamefont {Zocco}, \citenamefont {Wolf},
  \citenamefont {Merz}, \citenamefont {Schweiss}, \citenamefont {Heid},
  \citenamefont {Eder}, \citenamefont {Yu}, \citenamefont {Zhu}, \citenamefont
  {Si}, \citenamefont {Shibauchi},\ and\ \citenamefont
  {v.~L{\"o}hneysen}}]{eilers16}%
  \BibitemOpen
  \bibfield  {author} {\bibinfo {author} {\bibfnamefont {F.}~\bibnamefont
  {Eilers}}, \bibinfo {author} {\bibfnamefont {K.}~\bibnamefont {Grube}},
  \bibinfo {author} {\bibfnamefont {D.~A.}\ \bibnamefont {Zocco}}, \bibinfo
  {author} {\bibfnamefont {T.}~\bibnamefont {Wolf}}, \bibinfo {author}
  {\bibfnamefont {M.}~\bibnamefont {Merz}}, \bibinfo {author} {\bibfnamefont
  {P.}~\bibnamefont {Schweiss}}, \bibinfo {author} {\bibfnamefont
  {R.}~\bibnamefont {Heid}}, \bibinfo {author} {\bibfnamefont {R.}~\bibnamefont
  {Eder}}, \bibinfo {author} {\bibfnamefont {R.}~\bibnamefont {Yu}}, \bibinfo
  {author} {\bibfnamefont {J.-X.}\ \bibnamefont {Zhu}}, \bibinfo {author}
  {\bibfnamefont {Q.}~\bibnamefont {Si}}, \bibinfo {author} {\bibfnamefont
  {T.}~\bibnamefont {Shibauchi}},\ and\ \bibinfo {author} {\bibfnamefont
  {H.}~\bibnamefont {v.~L{\"o}hneysen}},\ }\href@noop {} {\bibfield  {journal}
  {\bibinfo  {journal} {Phys. Rev. Lett.}\ }\textbf {\bibinfo {volume} {116}},\
  \bibinfo {pages} {237003} (\bibinfo {year} {2016})}\BibitemShut {NoStop}%
\bibitem [{\citenamefont {Vanderkooy}\ and\ \citenamefont
  {Datars}(1968)}]{vanderkooy68}%
  \BibitemOpen
  \bibfield  {author} {\bibinfo {author} {\bibfnamefont {J.}~\bibnamefont
  {Vanderkooy}}\ and\ \bibinfo {author} {\bibfnamefont {W.~R.}\ \bibnamefont
  {Datars}},\ }\href@noop {} {\bibfield  {journal} {\bibinfo  {journal} {Can.
  J. Phys.}\ }\textbf {\bibinfo {volume} {46}},\ \bibinfo {pages} {1215}
  (\bibinfo {year} {1968})}\BibitemShut {NoStop}%
\bibitem [{\citenamefont {Barber}\ \emph {et~al.}(2018)\citenamefont {Barber},
  \citenamefont {Gibbs}, \citenamefont {Maeno}, \citenamefont {Mackenzie},\
  and\ \citenamefont {Hicks}}]{barber18}%
  \BibitemOpen
  \bibfield  {author} {\bibinfo {author} {\bibfnamefont {M.~E.}\ \bibnamefont
  {Barber}}, \bibinfo {author} {\bibfnamefont {A.~S.}\ \bibnamefont {Gibbs}},
  \bibinfo {author} {\bibfnamefont {Y.}~\bibnamefont {Maeno}}, \bibinfo
  {author} {\bibfnamefont {A.~P.}\ \bibnamefont {Mackenzie}},\ and\ \bibinfo
  {author} {\bibfnamefont {C.~W.}\ \bibnamefont {Hicks}},\ }\href@noop {}
  {\bibfield  {journal} {\bibinfo  {journal} {Phys. Rev. Lett.}\ }\textbf
  {\bibinfo {volume} {120}},\ \bibinfo {pages} {076602} (\bibinfo {year}
  {2018})}\BibitemShut {NoStop}%
\bibitem [{\citenamefont {Wu}\ \emph {et~al.}(2016)\citenamefont {Wu},
  \citenamefont {Zhao}, \citenamefont {Wang}, \citenamefont {Wang},
  \citenamefont {Xiang}, \citenamefont {Luo}, \citenamefont {Wu},\ and\
  \citenamefont {Chen}}]{wu16}%
  \BibitemOpen
  \bibfield  {author} {\bibinfo {author} {\bibfnamefont {Y.~P.}\ \bibnamefont
  {Wu}}, \bibinfo {author} {\bibfnamefont {D.}~\bibnamefont {Zhao}}, \bibinfo
  {author} {\bibfnamefont {A.~F.}\ \bibnamefont {Wang}}, \bibinfo {author}
  {\bibfnamefont {N.~Z.}\ \bibnamefont {Wang}}, \bibinfo {author}
  {\bibfnamefont {Z.~J.}\ \bibnamefont {Xiang}}, \bibinfo {author}
  {\bibfnamefont {X.~G.}\ \bibnamefont {Luo}}, \bibinfo {author} {\bibfnamefont
  {T.}~\bibnamefont {Wu}},\ and\ \bibinfo {author} {\bibfnamefont {X.~H.}\
  \bibnamefont {Chen}},\ }\href@noop {} {\bibfield  {journal} {\bibinfo
  {journal} {Phys. Rev. Lett.}\ }\textbf {\bibinfo {volume} {116}},\ \bibinfo
  {pages} {147001} (\bibinfo {year} {2016})}\BibitemShut {NoStop}%
\bibitem [{\citenamefont {Haule}\ and\ \citenamefont
  {Kotliar}(2009)}]{haule09}%
  \BibitemOpen
  \bibfield  {author} {\bibinfo {author} {\bibfnamefont {K.}~\bibnamefont
  {Haule}}\ and\ \bibinfo {author} {\bibfnamefont {G.}~\bibnamefont
  {Kotliar}},\ }\href@noop {} {\bibfield  {journal} {\bibinfo  {journal} {New
  J. Phys.}\ }\textbf {\bibinfo {volume} {11}},\ \bibinfo {pages} {025021}
  (\bibinfo {year} {2009})}\BibitemShut {NoStop}%
\bibitem [{\citenamefont {Yin}\ \emph {et~al.}(2011)\citenamefont {Yin},
  \citenamefont {Haule},\ and\ \citenamefont {Kotliar}}]{yin11}%
  \BibitemOpen
  \bibfield  {author} {\bibinfo {author} {\bibfnamefont {Z.~P.}\ \bibnamefont
  {Yin}}, \bibinfo {author} {\bibfnamefont {K.}~\bibnamefont {Haule}},\ and\
  \bibinfo {author} {\bibfnamefont {G.}~\bibnamefont {Kotliar}},\ }\href@noop
  {} {\bibfield  {journal} {\bibinfo  {journal} {Nat. Mater.}\ }\textbf
  {\bibinfo {volume} {10}},\ \bibinfo {pages} {932} (\bibinfo {year}
  {2011})}\BibitemShut {NoStop}%
\bibitem [{\citenamefont {{de' Medici}}\ \emph {et~al.}(2014)\citenamefont
  {{de' Medici}}, \citenamefont {Giovannetti},\ and\ \citenamefont
  {Capone}}]{demedici14}%
  \BibitemOpen
  \bibfield  {author} {\bibinfo {author} {\bibfnamefont {L.}~\bibnamefont {{de'
  Medici}}}, \bibinfo {author} {\bibfnamefont {G.}~\bibnamefont
  {Giovannetti}},\ and\ \bibinfo {author} {\bibfnamefont {M.}~\bibnamefont
  {Capone}},\ }\href@noop {} {\bibfield  {journal} {\bibinfo  {journal} {Phys.
  Rev. Lett.}\ }\textbf {\bibinfo {volume} {112}},\ \bibinfo {pages} {177001}
  (\bibinfo {year} {2014})}\BibitemShut {NoStop}%
\bibitem [{\citenamefont {Backes}\ \emph {et~al.}(2015)\citenamefont {Backes},
  \citenamefont {Jeschke},\ and\ \citenamefont {Valent{\'i}}}]{backes15}%
  \BibitemOpen
  \bibfield  {author} {\bibinfo {author} {\bibfnamefont {S.}~\bibnamefont
  {Backes}}, \bibinfo {author} {\bibfnamefont {H.~O.}\ \bibnamefont
  {Jeschke}},\ and\ \bibinfo {author} {\bibfnamefont {R.}~\bibnamefont
  {Valent{\'i}}},\ }\href@noop {} {\bibfield  {journal} {\bibinfo  {journal}
  {Phys. Rev. B}\ }\textbf {\bibinfo {volume} {92}},\ \bibinfo {pages} {195128}
  (\bibinfo {year} {2015})}\BibitemShut {NoStop}%
\bibitem [{\citenamefont {Zhao}\ \emph {et~al.}(2020)\citenamefont {Zhao},
  \citenamefont {Wo}, \citenamefont {Li}, \citenamefont {Song}, \citenamefont
  {Zheng}, \citenamefont {Li}, \citenamefont {Nie}, \citenamefont {Luo},
  \citenamefont {Zhao}, \citenamefont {Wu},\ and\ \citenamefont
  {Chen}}]{zhao20}%
  \BibitemOpen
  \bibfield  {author} {\bibinfo {author} {\bibfnamefont {D.}~\bibnamefont
  {Zhao}}, \bibinfo {author} {\bibfnamefont {H.~L.}\ \bibnamefont {Wo}},
  \bibinfo {author} {\bibfnamefont {J.}~\bibnamefont {Li}}, \bibinfo {author}
  {\bibfnamefont {D.~W.}\ \bibnamefont {Song}}, \bibinfo {author}
  {\bibfnamefont {L.~X.}\ \bibnamefont {Zheng}}, \bibinfo {author}
  {\bibfnamefont {S.~J.}\ \bibnamefont {Li}}, \bibinfo {author} {\bibfnamefont
  {L.~P.}\ \bibnamefont {Nie}}, \bibinfo {author} {\bibfnamefont {X.~G.}\
  \bibnamefont {Luo}}, \bibinfo {author} {\bibfnamefont {J.}~\bibnamefont
  {Zhao}}, \bibinfo {author} {\bibfnamefont {T.}~\bibnamefont {Wu}},\ and\
  \bibinfo {author} {\bibfnamefont {X.~H.}\ \bibnamefont {Chen}},\ }\href@noop
  {} {\bibfield  {journal} {\bibinfo  {journal} {Phys. Rev. B}\ }\textbf
  {\bibinfo {volume} {101}},\ \bibinfo {pages} {064511} (\bibinfo {year}
  {2020})}\BibitemShut {NoStop}%
\end{thebibliography}
%

\end{document}